\documentclass[preprint]{aastex61}

\usepackage{physics}

\newcommand\aastex{AAS\TeX}

\usepackage{bm}

\shorttitle{\aastex\ Calibration Errors in Radio Polarimetry}
\shortauthors{Hales}

\begin{document}

\title{Calibration Errors in Interferometric Radio Polarimetry}

\author[0000-0002-3733-2565]{Christopher A. Hales}
\email{chris.hales@newcastle.ac.uk}
\altaffiliation{Marie Sk\l{}odowska-Curie Fellow}
\affiliation{National Radio Astronomy Observatory, PO Box 0, Socorro, NM 87801, USA}
\affiliation{School of Mathematics and Statistics, Newcastle University, Newcastle upon Tyne NE1 7RU, UK}

\begin{abstract}

Residual calibration errors are difficult to predict in interferometric radio polarimetry
because they depend on the employed observational calibration strategy, encompassing the
Stokes vector of the calibrator and parallactic angle coverage. This work presents analytic
derivations and simulations that enable examination of residual on-axis instrumental leakage and
position angle errors for a suite of calibration strategies. The focus is on arrays comprising
alt-az antennas with common feeds over which parallactic angle is approximately uniform.
The results indicate that calibration schemes requiring parallactic angle coverage in the
linear feed basis (e.g. ALMA) need only observe over $30^\circ$, beyond which no significant
improvements in calibration accuracy are obtained. In the circular feed basis (e.g. VLA
above 1~GHz), $30^\circ$ is also appropriate when the Stokes vector of the leakage
calibrator is known a priori, but this rises to $90^\circ$ when the Stokes vector is unknown.
These findings illustrate and quantify concepts that were previously obscure rules of thumb.
\end{abstract}


\section{Introduction} \label{sec:introduction}

The mathematical foundations of interferometric radio polarimetry were first formulated
two decades ago by \citet{1996A&AS..117..137H}. Their work marked a significant
improvement over the previous `black box' approach presented by \citet{1964ApJ...139..551M}
(see also \citealt{1966ARA&A...4..245G} for an early history of radio polarimetry). The
formalism developed by \citet{1996A&AS..117..137H} continues to underpin cutting edge
developments in calibration \citep[e.g.][]{2011A&A...527A.106S}.

In a companion paper, \citet{1996A&AS..117..149S} examined the practicalities of
calibrating an array in the presence of errors due to incorrect assumptions about
calibrators or poorly determined parameters in the calibration process. An insightful
summary of the manner in which these various influences will lead to corruption of
calibration parameters was presented in Table~1 of \citet{1996A&AS..117..149S}.
However, it is not trivial to convert between the parameters presented in this
seminal work and the errors implied for calibrated data because this requires taking
into account the employed observational calibration strategy. An accessible overview
of these errors is needed to design efficient observing schemes in the present era
of increasing telescope automation. It is also needed more generally as an
educational tool to foster deeper understanding of radio polarimetry.

Some efforts have been made to improve this situation, such as the investigation
of dynamic range limitations presented by \citet{evlamemo177}. However, other
fundamental questions remain unaddressed. For example, what is the fractional
polarization below which a calibrator can be assumed to be unpolarized, or
the requirements on a calibrator's minimum coverage in parallactic
angle\footnote{Parallactic angle is that constructed at a celestial
coordinate between a line of constant right ascension and one pointing toward
zenith, as viewed from a geographic coordinate. It describes the orientation of the
sky as it rotates within the field of view of an alt-az telescope. To illustrate,
see \url{https://github.com/chrishales/plotparang} for publicly available code to
plot parallactic angle coverage while accounting for telescope elevation limits.}
throughout synthesis, to ensure that post-calibration residual instrumental polarization
or position angle errors remain below nominated thresholds.

The aim of this work is to fill the gap by presenting a practical guide to
errors in interferometric radio polarimetry that explicitly accounts for
calibration strategy. The structure is as follows. Section~\ref{sec:assumptions}
presents general assumptions concerning the types of interferometric arrays and
the observational parameter space that will be considered in this work.
Section~\ref{sec:polfund} presents a concise overview of polarization fundamentals,
providing conceptual and notational context for the remaining work.
Section~\ref{sec:limits} explores calibration strategies involving unpolarized
and polarized calibrators, examining the roles that a calibrator's fractional
polarization, signal to noise, and parallactic angle coverage play in limiting
post-calibration residual instrumental polarization. The same parameter space
is explored in Section~\ref{sec:pa} but with a focus on position
angle errors. Section~\ref{sec:conclusions} concludes.

\section{Assumptions}\label{sec:assumptions}

This work will assume an interferometric array with the following characteristics.
First, all feeds are alt-az mounted, i.e. situated on alt-az antennas without
dish rotators such as those used on the Australian Square Kilometre Array Pathfinder (ASKAP).
Second, there are at least 3 baselines; closure between $N\ge3$ antennas is required to
solve for of order $N$ matrices from $N(N-1)/2$ baselines \citep{1996A&AS..117..149S}.
Third, parallactic angle ($\psi$) is constant across the array, within mechanical alignment
errors, such that $\psi_i=\psi$ for all $i$ antennas (i.e. the array is small).
And fourth, all antennas are fitted with dual orthogonal linear (X, Y) or circular (R, L)
feeds with the same nominal alignment, forming a homogeneous array. From these, all
polarization products are measured.

Polarization calibrators will be assumed unresolved and located on-axis.
The off-axis polarimetric response of the system will not be considered.
Any frequency dependence of parameters will be assumed implicitly.

This work will focus on linear polarimetry\footnote{Circular, linear, and elliptical
polarization will be indicated in this work by $\mathcal{V}$,
$\mathcal{L}=\sqrt{\mathcal{Q}^2+\mathcal{U}^2}$, and
$\mathcal{P}=\sqrt{\mathcal{V}^2+\mathcal{L}^2}$, respectively, formed from
the Stokes parameters $\mathcal{Q}$, $\mathcal{U}$, and $\mathcal{V}$. Fractional
values are divided by Stokes $\mathcal{I}$.}.
Circular polarization calibration will be touched on for completeness in
Section~\ref{sec:polfund}, but no error analysis will be pursued (where circular
to linear leakage effects, or vice versa, may be important).

Results will be illustrated for two representative arrays: the 
Atacama Large Millimeter/submillimeter Array (ALMA) which observes with
linear feeds in all bands, and the Karl G. Jansky Very Large Array (VLA)
which observes with circular feeds in all bands above 1~GHz.

Terminology and notation throughout this work will follow conventions
from the widely used Common Astronomy Software Applications package
\citep[CASA;][]{2007ASPC..376..127M}.

\section{Polarization Fundamentals}\label{sec:polfund}

The radio interferometer measurement equation
\citep{1996A&AS..117..137H,1996A&AS..117..149S,noordam,2011A&A...527A.106S} relates observed
visibilities to ideal model visibilities on a baseline between antennas $i$ and $j$ as
\begin{equation}
    \bm{V^{obs}}_{ij} = \mathbf{B}_{ij}\,\mathbf{G}_{ij}\,\mathbf{D}_{ij}\,\mathbf{P}_{ij}\,
                        \bm{V^{mod}}_{ij} \label{eqn:me}
\end{equation}
where the corrupting Mueller matrix terms (frequency-dependent outer products of antenna-based
Jones matrices; $\mathbf{M}_{ij}=\mathbf{J}_i\otimes\mathbf{J}_j^*$) are associated from right
to left with parallactic angle, instrumental polarization leakage, combined electronic
and atmospheric gains, and bandpass, respectively. Some terms are neglected above for clarity
(e.g. Faraday rotation and phase delay associated with the ionosphere/plasmasphere, antenna
elevation-dependence, non antenna-based terms). Calibration involves solving for these terms
and applying their inverse to the observed data to recover corrected data
\begin{equation}
    \bm{V^{corr}} = \mathbf{P}^{-1}\,\mathbf{D}^{-1}\,\mathbf{G}^{-1}\,
                    \mathbf{B}^{-1}\,\bm{V^{obs}} \,
\end{equation}
where antenna indices will now be omitted unless required for clarity. While the corrupting
terms in the measurement equation are written as independent effects along
the signal path (i.e. from right to left in Equation~\ref{eqn:me}), in general
they are not. Care must therefore be taken to distinguish dominant terms from those that are coupled
to others. The former can be solved for independently, while the latter require an iterative approach
to converge on a global solution over multiple terms\footnote{To illustrate, see the CASA approach to
calibration described at \url{https://casa.nrao.edu/docs/UserMan/casa_cookbook016.html}.}.

The measurement equation is typically refactored to the relative phase frame of the bandpass/gain
reference antenna (phase fixed to zero in both polarizations on this antenna)
\begin{equation}
    \bm{V^{obs}} = \mathbf{B_r}\,\mathbf{G_r}\,\mathbf{K_{crs}}\,\mathbf{D_r}\,
                       \mathbf{\widetilde{X}_r}\,\mathbf{P}\,\bm{V^{mod}}
\end{equation}
in which the refactored terms are identified by subscript r, the crosshand bandpass
phase\footnote{Also known in the literature as the $XY$ phase, $RL$ phase, or phase-zero difference.}
on the reference antenna $\mathbf{X_r}$ that arises from the refactoring
of $\mathbf{B}\,\mathbf{G} = \mathbf{B_r}\,\mathbf{G_r}\,\mathbf{X_r}$ remains unconstrained,
$\mathbf{X_r}=\mathbf{K_{crs}}\,\mathbf{\widetilde{X}_r}$ is separated into crosshand delay
$\mathbf{K_{crs}}$ and crosshand phase $\mathbf{\widetilde{X}_r}$ terms that capture
first-order linear and residual non-linear frequency dependence, respectively, and leakages
$\mathbf{D_r}=\mathbf{X_r}\,\mathbf{D}\,\mathbf{X^{-1}_r}$ are measured in the crosshand phase
frame\footnote{Note that if polarization calibration will be performed, the same reference
antenna must be used for all calibration solutions. If this condition is not met, the crosshand phase
frame will be ambiguous and polarization calibration will be corrupted. This is not a requirement
when calibrating only parallel-hand visibility data, which are insensitive to crosshand phase.}.

The leakage terms (`dipole' terms or d-terms) describe imperfections in the polarimetric response
of the system and quantify the degree to which each feed is sensitive to an orthogonally
polarized signal. The imperfections arise from both telescope geometry (e.g.
antenna illumination, feed horn, optics alignment) and electronic hardware (e.g. polarization
splitter, hybrid coupler). Notation for leakages in this work will follow the Jones matrix form
\begin{equation}
    \mathbf{D}_i =
    \begin{bmatrix}
        1 & d_{pi}(\nu) \\
        d_{qi}(\nu) & 1
    \end{bmatrix}
\end{equation}
for antenna $i$ where $p$ is given by $X$ (linear basis) or $R$ (circular basis), $d_{pi}$ is
the fraction of the $q$ (orthogonal) polarization sensed by $p$, and on-diagonal effects are factored
into $\mathbf{B}$ and $\mathbf{G}$ (\citealt{1996A&AS..117..149S}\footnote{Note that
\citet{1996A&AS..117..149S} use a different sign convention for d-terms.}). Frequency dependence
($\nu$) will be assumed implicitly throughout this work and omitted below. Other terms from the
measurement equation that are most relevant to this work are parallactic angle in the linear
feed (LF) and circular feed (CF) bases and crosshand phase on the reference antenna. These take
the respective Jones forms
\begin{equation}
    \mathbf{P}^{^\textrm{\tiny LF}}_i =
    \begin{bmatrix}
        \cos{\psi} & \sin{\psi} \\
        -\sin{\psi} & \cos{\psi}
    \end{bmatrix}
\end{equation}
and
\begin{equation}
    \mathbf{P}^{^\textrm{\tiny CF}}_i =
    \begin{bmatrix}
        e^{-i\psi} & 0 \\
        0 & e^{i\psi}
    \end{bmatrix}
\end{equation}
where $\psi$ is parallactic angle, and
\begin{equation}
    \mathbf{X_r} =
    \begin{bmatrix}
        e^{i\rho} & 0 \\
        0 & 1
    \end{bmatrix}
\end{equation}
where $\rho$ is crosshand phase.

For an interferometer with dual linearly polarized feeds, $\bm{V^{mod}}$ is given by the 4-element vector
\begin{eqnarray}
    V_{XX} &=& \mathcal{I} + \mathcal{Q} \\
    V_{XY} &=& \mathcal{U} + i\,\mathcal{V} \\
    V_{YX} &=& \mathcal{U} - i\,\mathcal{V} \\
    V_{YY} &=& \mathcal{I} - \mathcal{Q} \;\;,
\end{eqnarray}
whereas for circular feeds the vector is
\begin{eqnarray}
    V_{RR} &=& \mathcal{I} + \mathcal{V} \\
    V_{RL} &=& \mathcal{Q} + i\,\mathcal{U} \\
    V_{LR} &=& \mathcal{Q} - i\,\mathcal{U} \\
    V_{LL} &=& \mathcal{I} - \mathcal{V} \;\;.
\end{eqnarray}
The model visibilities for a single baseline, corrupted by parallactic angle,
leakage, and crosshand phase terms
($\mathbf{X_r}\,\mathbf{D}\,\mathbf{P}\,\bm{V^{mod}} = \mathbf{G}^{-1}\,\mathbf{B}^{-1}\,\bm{V^{obs}}$),
are given in the linear basis by
\begin{eqnarray}
    V_{XX} &=&                           (\mathcal{I}      +    \mathcal{Q_\psi}) +
               \mathcal{U_\psi}       \, (d_{Xi} + d_{Xj}^{\,*}) \label{eqn:vxx}\\
    V_{XY} &=&                     \big[ (\mathcal{U_\psi} + i\,\mathcal{V}) +
               \mathcal{I}            \, (d_{Xi} + d_{Yj}^{\,*}) -
               \mathcal{Q_\psi}       \, (d_{Xi} - d_{Yj}^{\,*}) \big] \, e^{i\rho} \label{eqn:vxy}\\
    V_{YX} &=&                     \big[ (\mathcal{U_\psi} - i\,\mathcal{V}) +
               \mathcal{I}            \, (d_{Yi} + d_{Xj}^{\,*}) +
               \mathcal{Q_\psi}       \, (d_{Yi} - d_{Xj}^{\,*}) \big] \, e^{-i\rho} \label{eqn:vyx}\\
    V_{YY} &=&                           (\mathcal{I}      -    \mathcal{Q_\psi}) +
               \mathcal{U_\psi}       \, (d_{Yi} + d_{Yj}^{\,*}) \label{eqn:vyy}
\end{eqnarray}
where $\mathcal{Q_\psi} \equiv \mathcal{Q}\cos{2\psi} + \mathcal{U}\sin{2\psi} $,
$\mathcal{U_\psi} \equiv \mathcal{U}\cos{2\psi} - \mathcal{Q}\sin{2\psi} $,
and terms multiplied by second order leakages (e.g. $d_{Xi}\,d_{Xj}^{\,*}$) are neglected.
The visibilities in the circular basis are given by
\begin{eqnarray}
    V_{RR} &=&                           (\mathcal{I} +    \mathcal{V})            +
               d_{Ri}                 \, (\mathcal{Q} - i\,\mathcal{U})e^{+i2\psi} +
               d_{Rj}^{\,*}           \, (\mathcal{Q} + i\,\mathcal{U})e^{-i2\psi} \\
    V_{RL} &=&                     \big[ (\mathcal{Q} + i\,\mathcal{U})e^{-i2\psi} +
               \mathcal{I}            \, (d_{Ri} + d_{Lj}^{\,*})                   +
	       \mathcal{V}            \, (d_{Lj}^{\,*} - d_{Ri})                   \big] e^{i\rho} \label{eqn:vrl} \\
    V_{LR} &=&                     \big[ (\mathcal{Q} - i\,\mathcal{U})e^{+i2\psi} +
               \mathcal{I}            \, (d_{Li} + d_{Rj}^{\,*})                   +
               \mathcal{V}            \, (d_{Li} - d_{Rj}^{\,*})                   \big] e^{-i\rho} \\
    V_{LL} &=&                           (\mathcal{I} -    \mathcal{V})            +
              d_{Li}                 \, (\mathcal{Q} + i\,\mathcal{U})e^{-i2\psi} +
              d_{Lj}^{\,*}           \, (\mathcal{Q} - i\,\mathcal{U})e^{+i2\psi} \;\;.
\end{eqnarray}

Analysis in this work will be restricted to the linearized equations presented above
(first order in d-terms). Furthermore, polarization calibration will be limited to
examination of cross hand visibilities only, with the additional assumption that
Stokes $\mathcal{V}$ is zero (unless a non-zero model is available). These simplifications
are currently used within CASA and are suitable for the focus on linear polarization
in this work \citep[note that the full quadratic approach is used in other packages such
as MIRIAD;][]{1995ASPC...77..433S}.

Crosshand phase calibration can be performed (e.g. in CASA) by taking the average
over all baselines for the sum of cross hand visibilities,
\begin{eqnarray}
    \left<V_{XY}\right>+\left<V_{YX}^{\,*}\right> &=&
    e^{i\rho} \big[
    \mathcal{U_\psi} + 
    \mathcal{I} \left<
    d_{Xi} + d_{Yi}^{\,*} + d_{Xj} + d_{Yj}^{\,*} \right> - \nonumber\label{eqn:chp:lin} \\
    && \mathcal{Q_\psi} \left<
    d_{Xi} - d_{Yi}^{\,*} + d_{Xj} - d_{Yj}^{\,*} \right>
    \big] \label{eqn:xl} \\
    \left<V_{RL}\right>+\left<V_{LR}^{\,*}\right> &=&
    e^{i\rho} \big[
    \left(\mathcal{Q} + i\,\mathcal{U}\right)e^{-i2\psi} +
    \mathcal{I} \left<
    d_{Ri} + d_{Li}^{\,*} + d_{Rj} + d_{Lj}^{\,*}
    \right> \big] \;\;. \label{eqn:rlphase}
\end{eqnarray}
In the circular basis, when leakages are known, crosshand phase calibration is synonymous with
calibration of the absolute alignment of linear polarization\footnote{This is only strictly true
for infinite signal to noise. In practice there will be a (likely) negligible yet non-zero bias
between the recovered crosshand phase and the true overall position angle correction needed to
correctly orient the crosshand phase frame.} and requires an external source of known position angle.
In the linear basis, an offset in the absolute alignment of the feeds (i.e. different observed
$\mathcal{U_\psi}$ and $\mathcal{Q_\psi}$ in Equation~\ref{eqn:xl}) does not translate into a trivial
change in crosshand phase. Thus, in the linear basis, crosshand phase and absolute position angle
calibrations are not synonymous. However, unlike in the circular basis, if the linear antenna
feeds are nominally aligned to the sky, an external source of known position angle is not
formally required; variation in $\mathcal{U_\psi}$ over parallactic angle for a linearly
polarized source with unknown $\mathcal{Q}$ and $\mathcal{U}$ is sufficient to solve for $\rho$.
As a result, calibration strategies in the linear basis typically need to obtain a first-pass
solution for $\mathbf{X_r}$, assuming zero leakages, prior to solving for $\mathbf{D_r}$.
Subsequent iteration is technically required (though typically negligible in practice)
to account for leakages in the $\mathbf{X_r}$ solve. In the
circular basis, $\mathbf{X_r}$ is not needed to solve for the leakages (crosshand
phase simply imparts an overall rotational ambiguity) while $\mathbf{D_r}$ is needed
to optimally solve for $\mathbf{X_r}$. Thus circular basis calibration strategies
typically solve for $\mathbf{D_r}$ prior to $\mathbf{X_r}$.

\begin{figure}
\centerline{\includegraphics[clip,width=0.5\textwidth]{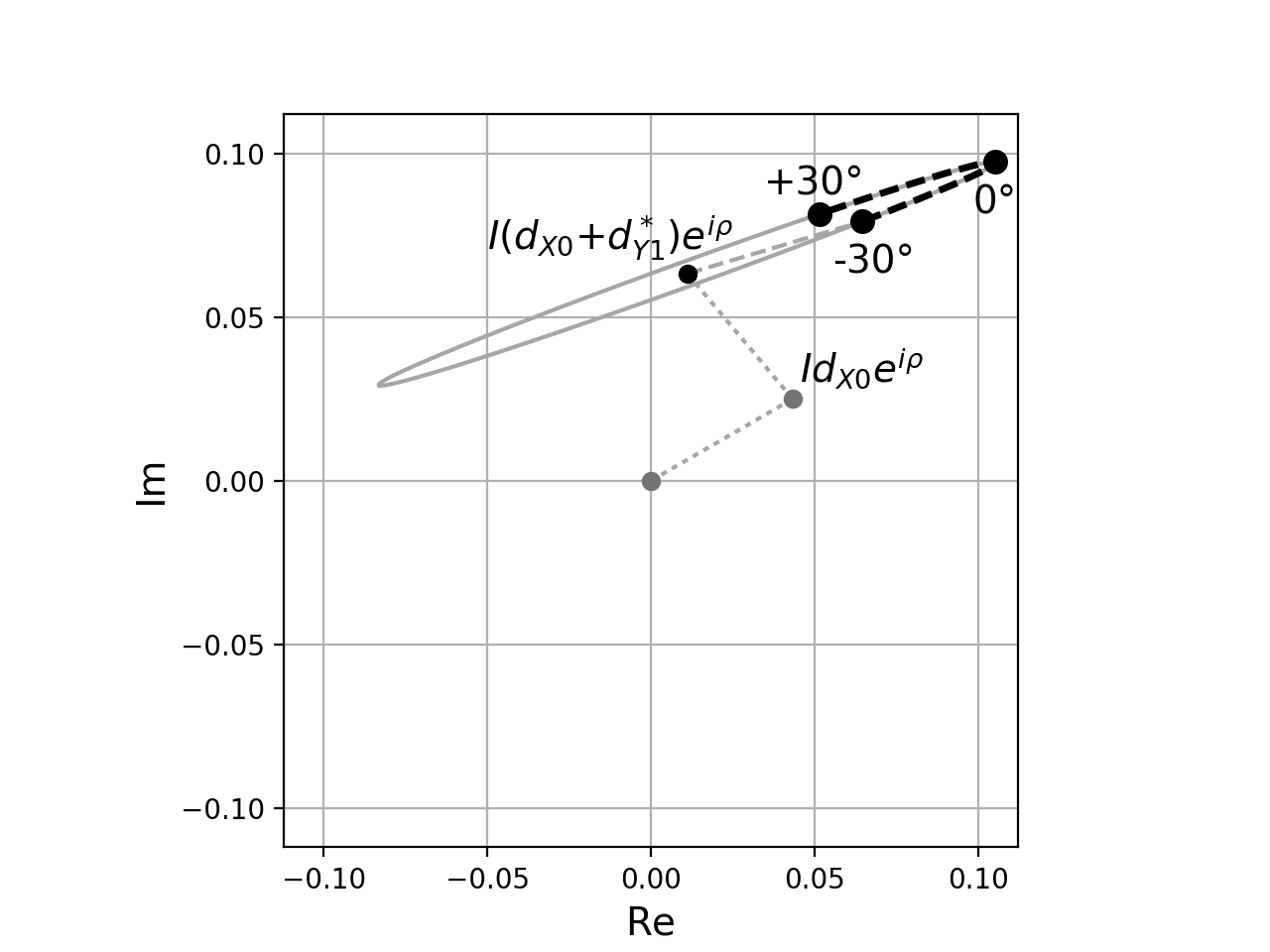}}
\caption{
Path traced in complex plane by corrupted $V_{XY}$ from Equation~\ref{eqn:vxy}
on a single baseline ($i=0$, $j=1$) for a calibrator with
$(\mathcal{I},\mathcal{Q},\mathcal{U},\mathcal{V})=(1,0,0.1,0)$, instrumental
leakages $d_{X0}$ and $d_{Y1}$ with moduli 0.05 and respective phases $10^\circ$
and $-110^\circ$, and crosshand phase $\rho=20^\circ$. The ellipse indicates the
path traced by complete parallactic angle coverage, while the dashed curve indicates
the path traced from $\psi=-30^\circ$ to $+30^\circ$ through $0^\circ$. The
lighter-shaded dashed line connects the center of the ellipse to $\psi=-30^\circ$.
The lighter-shaded dotted lines and points indicate how leakage from total intensity
offsets the center of the ellipse from zero.
}
\label{fig:geomXY}
\end{figure}

\begin{figure}
\centerline{\includegraphics[clip,width=0.5\textwidth]{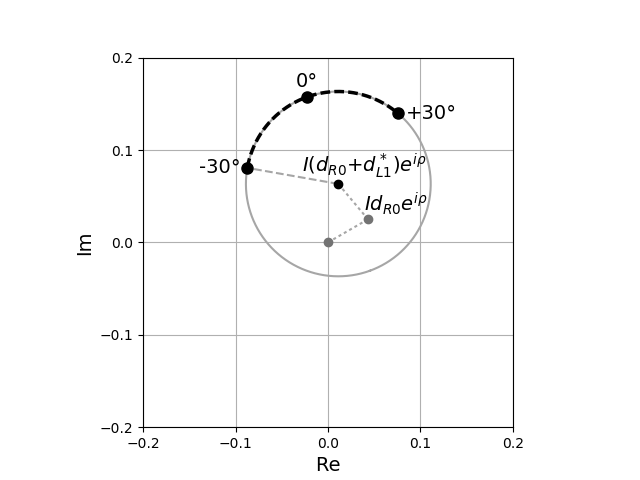}}
\caption{
Path traced in complex plane by corrupted $V_{RL}$ from Equation~\ref{eqn:vrl}
for a single baseline. All calibrator and instrumental parameters are the same
as described for Fig.~\ref{fig:geomXY}, but with leakage subscripts $X$ and $Y$
replaced by $R$ and $L$, respectively. Full parallactic angle coverage traces
a circle, offset from zero by leakage from total intensity.
}
\label{fig:geomRL}
\end{figure}

Figures~\ref{fig:geomXY} and \ref{fig:geomRL} illustrate how the respective
corrupted cross hands from Equations~\ref{eqn:vxy} and \ref{eqn:vrl} trace
out geometric features in the complex plane as a function of parallactic angle coverage.
In the linear basis, $\mathcal{U_\psi}$ moves along a rotated linear axis
that is offset from the origin by leakages and broadened to become an
ellipse due to the product of leakages with $\mathcal{Q_\psi}$. In the
circular basis, a circle is drawn with center offset by the leakages
\citep[e.g.][]{1969MNRAS.142...11C}. These
figures highlight degrees of freedom that must be calibrated. For example, calibration
strategies in the circular basis that involve a polarized calibrator with unknown
Stokes vector require at least 3 independent observations to solve for the d-terms
(provided there are at least 3 baselines) as well as Stokes $\mathcal{Q}$ and
$\mathcal{U}$. Geometrically, this can be viewed as the need for 3 points to solve
for the unknown origin and radius of a circle. When the Stokes vector is known a
priori, only two observations are required to locate the origin (the known sense
of rotation between the observations breaks the origin degeneracy). Degrees of
freedom are examined more formally in the following section.

\subsection{Degrees of freedom}\label{sec:polfunddof}

Polarimetric calibration involves solving for the crosshand phase, leakage d-terms, and
absolute alignment of linear polarization. External calibration is required to determine
the absolute position angle in the same way that an interferometer cannot self-calibrate
the absolute flux density level. Theoretically, to solve for all degrees of freedom in
any basis that uses dual orthogonally polarized feeds, at least 3 distinct observations of
calibrators with linearly independent Stokes vectors are required \citep{1996A&AS..117..149S}.
This implies that at least 2 observations need to be on a polarized calibrator, at least 1
needs to be linearly polarized, and a circularly polarized calibrator is not essential.
Observation of a linearly polarized calibrator over a range of parallactic angles can
provide the necessary 3 distinct observations; rotation of the sky within the alt-az
instrument frame enables the leakages and source polarization to be jointly solved.
In practice, for circular polarization science in the linear feed basis, external (absolute)
calibration of Stokes $\mathcal{V}$ is also required\footnote{To avoid the need for a circularly
polarized calibrator in accord with the theoretical requirements presented above,
second-order d-terms must be taken into account to break the imaginary-axis degeneracy
otherwise present in the linearized $\mathbf{D}\,\mathbf{P}\,\bm{V^{mod}}$ equations.
However, even if these terms are included, non-singular solutions are likely to be produced
in practice (small leakages, thermal noise, gain stability), in turn requiring absolute
circular polarization calibration in the linear basis. To intuit why a similar issue
does not arise in the circular basis, note that in the limit of large leakages there
is no difference between observations in either basis, i.e. circular feeds can be
thought of as linear feeds with high leakages (or leakages that act with crosshand phase
to effectively operate as a quadrature hybrid coupler). In this case, the additional
constraints available through the linear basis second-order terms become accessible.}
\citep[e.g.][]{2000MNRAS.319..484R} due to leakages being small (as a result of good engineering).

\subsection{Absolute vs. relative leakages}\label{sec:polfundavr}

Observational constraints may not always be available to solve for the d-terms
unambiguously. For example, an unpolarized calibrator will yield solutions that are
degenerate in the sum of leakage pairs, e.g. $d_{Ri} + d_{Lj}^{\,*}$. This
corresponds to an undetermined Jones matrix which, in the small angle approximation,
corresponds to a complex offset ($\beta$) that can be added to one polarization and
subtracted with conjugation from the other; e.g. $d_{pi}^\prime = d_{pi} + \beta$ and
$d_{qj}^\prime = d_{qj} - \beta^{\,*}$ for all antennas \citep{1996A&AS..117..149S}.
When solving for degenerate leakages, the real and imaginary components of the
X or R feed leakages will be (arbitrarily) set to a constant (typically zero) on
the gain reference antenna, effectively setting $\beta$ to the negative of the true
d-term on this antenna (e.g. $\beta = -d_{p,ref}$). Leakage solutions degenerate
in this manner are known as relative leakages. Absolute leakages are only accessible
when additional observational constraints are available, e.g. from multiple
observations of a polarized calibrator.

Relative leakages cannot substitute absolute leakages in the measurement equation
without incurring errors. For calibration strategies that recover relative leakages,
the degenerate nature of the solutions will manifest in the linear basis as an error
in the position angle of linear polarization, an unknown degree of leakage between
linearly and circularly polarized components, and gain errors in total intensity.
In the circular basis there will be an unknown degree of leakage between linearly and
circularly polarized components and gain errors will be produced in total intensity
\citep[][or glean from equations presented in Section~\ref{sec:polfund}]{1996A&AS..117..149S}.
As described earlier, this work will only focus on leakages accessible through the
linearized cross hand visibilities. Accordingly, calibration strategies in the linear feed basis
may recover relative or absolute leakages, depending on available observational
constraints. Absolute leakages are accessible because the d-term sum degeneracy
can be broken purely in the cross hands by $\mathcal{Q_\psi}$. Leakages in the
circular basis, however, will always be relative; absolute leakages cannot be
easily recovered without accessing the quadratic terms in the parallel hand visibilities
or performing observations in which a subset of receivers are physically rotated
\citep{evlamemo170}.

For completeness, it is worth noting that absolute leakages will often be quasi-absolute
in practice because the measurement equation formalism assumes that the full signal path
is characterized by a specific and limited set of effects. For example, the measurement
equation under consideration may neglect terms regarding telescope analogue components
\citep{2015MNRAS.449..107P} or direction dependent effects \citep{2011A&A...527A.106S}.
In practice, absolute leakages are non-singular solutions within the assumed framework.

\section{Residual on-axis instrumental polarization}\label{sec:limits}

Strategies to calibrate instrumental leakage typically involve a single observation
of an unpolarized calibrator, or multiple observations of a polarized calibrator
spanning a range of parallactic angles. This section will present results from
analytic derivations and Monte Carlo simulations with the aim to elucidate
calibrator requirements so that subsequent observation of an unpolarized science
target will deliver spurious on-axis polarization below a nominated threshold.
For example, ALMA specifications require residual instrumental on-axis polarization
to be below 0.1\% of total intensity after calibration. Sections~\ref{sec:limits:unpol}
and \ref{sec:limits:pol} will examine unpolarized and polarized calibrators,
respectively, in both the linear and circular feed bases.

Throughout the following, an observation of a calibrator at a particular parallactic
angle will be termed a {\it slice}. A slice may comprise one or more observational
scans (in VLA parlance), but it will be assumed that parallactic angle is approximately
constant throughout the slice and that the quoted signal to noise represents all
combined scans within the slice. Note that, in practice, these concepts are linked:
the ability to define the timespan over which parallactic angle can be considered
constant is a function of signal to noise. Separation of these concepts is useful
for framing the simulations. However, caution is required to ensure that the
time needed to obtain a requisite signal to noise is not comparable to the
parallactic angle range over which significant changes in residual leakage
are predicted to occur.

A requirement for maximum spurious on-axis polarization translates to
a calibration requirement for d-term accuracy. Taking $\sigma_d$ as the
characteristic d-term modulus error\footnote{If characteristic errors
in either the real or imaginary d-term components are $\sigma$, then under Rayleigh
statistics $\sigma_d=\sqrt{\pi/2}\,\sigma$.} and $N_a$ as the number of antennas
in the array, the approximate level of spurious on-axis
linear ($\mathcal{L}_\epsilon^\textrm{\tiny LF}$) or
circular ($\mathcal{V}_\epsilon^\textrm{\tiny LF}$) polarization produced
when observing an unpolarized source in the linear feed basis is
\begin{equation}
  \mathcal{L}_\epsilon^\textrm{\tiny LF} \approx \mathcal{V}_\epsilon^\textrm{\tiny LF}
  \approx \mathcal{I}\frac{\sigma_d}{\sqrt{N_a}} \;\;.
  \label{eqn:residL:LF}
\end{equation}
Spurious elliptical polarization is
\begin{equation}
  \mathcal{P}_\epsilon^\textrm{\tiny LF}
  \approx \mathcal{I}\sigma_d\,\sqrt{\frac{\pi}{2 N_a}} \;\;.
\end{equation}
In the circular feed basis, the level of spurious linear or elliptical
polarization is
\begin{equation}
  \mathcal{L}_\epsilon^\textrm{\tiny CF} \approx \mathcal{P}_\epsilon^\textrm{\tiny CF}
  \approx \mathcal{I}\sigma_d\,\sqrt{\frac{\pi}{2 N_a}} \;\;.
  \label{eqn:residL:CF}
\end{equation}
No spurious circular polarization will be produced. Analytic derivations of these
equations are presented in the Appendix. To illustrate, a requirement of 0.1\%
spurious on-axis linear polarization translates to $\sigma_d$~{\footnotesize $\lesssim$}~0.6\%
for ALMA ($N_a=40$) and $\sigma_d$~{\footnotesize $\lesssim$}~0.4\% for the VLA ($N_a=27$).
The equations above assume a worst-case scenario where the science target is observed
within a single parallactic angle slice. For wider parallactic angle coverage the
residual leakage will be smaller due to depolarization.

The equations above will now be used to translate $\sigma_d$ into limits on
anticipated spurious on-axis polarization for various calibration strategies.

\subsection{Unpolarized calibrators}\label{sec:limits:unpol}

A calibrator that is classified as unpolarized may in fact exhibit a low level
of polarization, denoted by $\mathcal{L_\textrm{true}}$,
$\mathcal{U_{\psi,\textrm{true}}}$, or $\mathcal{V_\textrm{true}}$ (other
terms not needed below). Taking this into account, if leakage calibration is
performed using an assumed unpolarized calibrator (resulting in relative leakages),
the resulting d-term modulus error $\sigma_d$ will be approximately
\begin{equation}
  2 \left( \sigma_d^\textrm{\tiny LF} \right)^2 \approx
    \left(\frac{\mathcal{U_{\psi,\textrm{true}}}}{\mathcal{I}}\right)^{\!2} +
    \left(\frac{\mathcal{V_\textrm{true}}}{\mathcal{I}}\right)^{\!2} + 
    \frac{N_a}{A^2}
  \label{eqn:residpolLF}
\end{equation}
in the linear feed basis and
\begin{equation}
  2 \left( \sigma_d^\textrm{\tiny CF} \right)^2 \approx
    \left(\frac{\mathcal{L_\textrm{true}}}{\mathcal{I}}\right)^{\!2} +
    \frac{N_a}{A^2}
  \label{eqn:residpolCF}
\end{equation}
in the circular feed basis, where $A$ is the full-array dual-polarization total
intensity signal to noise of the calibrator within the single spectral channel of
interest. Derivations of these equations are presented in the Appendix. Note
that the level of fractional polarization below which a source may be classified
as `unpolarized' depends on the telescope being used and the science goals of
the observation. For science projects or telescopes that place a requirement on
the acceptable level of spurious on-axis polarization following calibration,
the equations above can be used to determine if a calibrator can be classed
as unpolarized. Note that the dynamic range limitations arising from calibrator
model errors presented by \citet{evlamemo177} may be relevant or even
dominate considerations for some science goals.

Leakage calibration with an unpolarized calibrator can be performed using a single
slice observation. In the circular basis, taking an example
of an assumed unpolarized calibrator with true linear polarization $\sim1\%$ observed
with the VLA at high signal to noise, the estimate from Equation~\ref{eqn:residpolCF}
is $\sigma_d\sim0.7\%$. The estimated spurious on-axis fractional polarization for an
unpolarized science target observed over a small range in parallactic angle (e.g. snapshot)
is then $\sim0.2\%$ from Equation~\ref{eqn:residL:CF}. As another example, now in the
linear basis using an unpolarized (or negligibly polarized) calibrator, the estimated
spurious fractional linear polarization for an unpolarized science target is $\sim1/\sqrt{2A^2}$.

\subsection{Polarized calibrators}\label{sec:limits:pol}

\subsubsection{Linear basis}\label{sec:limits:pol:lin}

To calibrate leakages in the linear basis using a polarized source, observations
are required over at least 3 parallactic angle slices if the Stokes vector is
unknown a priori, or as little as a single slice if the Stokes vector is known.
When the Stokes vector is unknown a priori, it needs to be solved for in addition to
the d-terms and crosshand phase. For the case of a single slice observation on a
calibrator with known Stokes vector, the available degrees of freedom only permit
solving for relative leakages. In all other calibration strategies, absolute
leakages can be recovered.

Simulations were performed to predict the level of spurious on-axis polarization and
absolute position angle error resulting from a 1 (linearly polarized with Stokes known),
2 (linearly polarized with Stokes known), 3 (Stokes unknown), and 10 (Stokes unknown)
slice strategy. Full details are presented in the Appendix, including a web link to
the simulation code which has been made publicly available. This section will present
the results for spurious polarization. Position angle errors will be reported in
Section~\ref{sec:pa:lin}.

The simulations explore the accuracy with which the d-terms, crosshand phase, and
calibrator polarization (when relevant) can be measured from the cross hand
visibilities when a source is subjected to parallactic angle rotation in the presence
of thermal noise. To solve for $\sigma_d$, the code focuses on a single cross product
($V_{XY}$) and examines how well the d-term for the antenna and polarization under
consideration can be recovered while taking into account all available $N_a-1$ baselines
toward antenna $i$. The simulations assume an ALMA-like array with 40 antennas,
typical d-term modulus of 1.5\%, and mechanical feed alignment uncertainty of $2^\circ$
per antenna \citep{cortesALMApol}; the alignment uncertainty is only used for position
angle analysis (see Section~\ref{sec:pa:lin}).

For multi-slice strategies, the first slice is assumed to be observed at maximum
$|\mathcal{U_\psi}|$. This produces approximately worst-case results, compared to for
example symmetric coverage about maximum $|\mathcal{U_\psi}|$, because the arc traced
along the ellipse in the complex plane for a cross hand visibility is minimized,
leading to poorer constraints on recovered parameters. Note that truly worst-case
results would arise from limited coverage about $\mathcal{U_\psi}=0$, in which case
crosshand phase would be degenerate and calibration would fail. The simulations use
Monte Carlo sampling to recover the distribution of $\sigma_d$, from which the
$95^{th}$ percentile is measured and converted to spurious linear polarization using
Equation~\ref{eqn:residL:LF}. Two scenarios have been examined, the first for a
calibrator exhibiting 3\% fractional linear polarization and the second with fraction 10\%.

\begin{figure}
\centerline{\includegraphics[clip,width=0.5\textwidth]{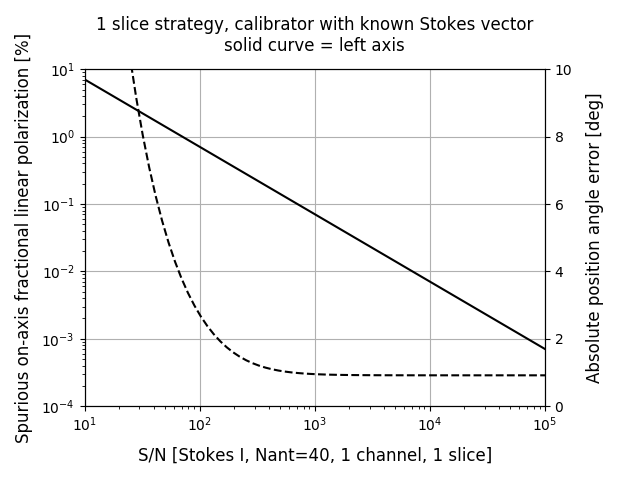}}
\caption{
Results from simulations showing $95^{th}$ percentile spurious on-axis fractional linear
polarization and absolute position angle error, predicted to arise within 1 spectral
channel when a linear basis telescope is calibrated using a single slice
observation of a polarized calibrator with known Stokes vector.
The indicated position angle error must be added in quadrature with a target
source's statistical error to obtain its total position angle error.
The displayed curves are from simulations with a 3\% linearly polarized
calibrator. Results from the 10\% case are not displayed as they are indistinguishable.
}
\label{fig:1sliceknown}
\end{figure}

Results for the 1 slice strategy are presented in Figure~\ref{fig:1sliceknown}.
The displayed spurious fractional polarization trend is practically indistinguishable
from the analytic prediction for an unpolarized calibrator described at the end
of Section~\ref{sec:limits:unpol}. The reason is because d-term errors arising from
crosshand phase errors, drawn from data where all baselines are combined in the solve,
always remain practically negligible compared to thermal noise in the subset
of baselines from which individual d-terms are effectively solved. Thus the fractional
polarization of the calibrator will not practically affect the result (unless
it approaches 100\%); curves obtained for the 3\% and 10\% fractionally polarized cases
are indistinguishable.

\begin{figure}[p]
\centerline{
\includegraphics[clip,width=0.5\textwidth]{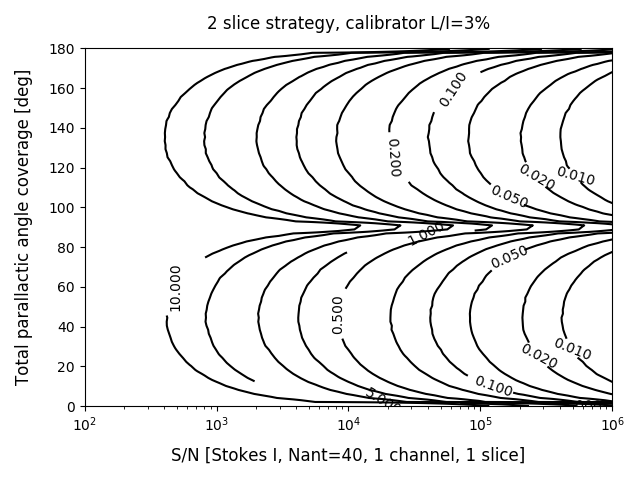}
\includegraphics[clip,width=0.5\textwidth]{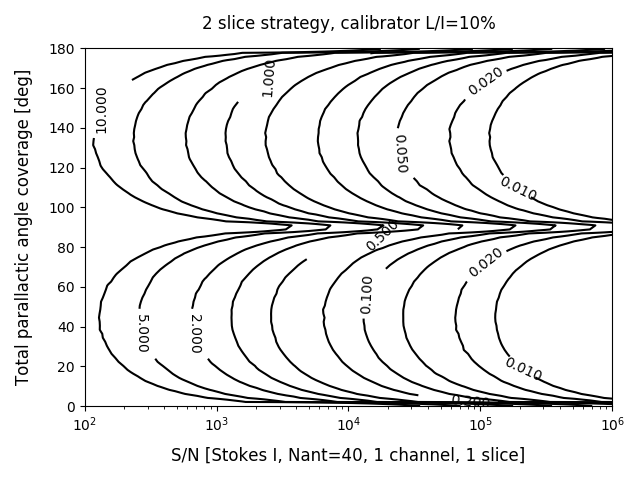}
}
\vspace{0mm}
\centerline{
\includegraphics[clip,width=0.5\textwidth]{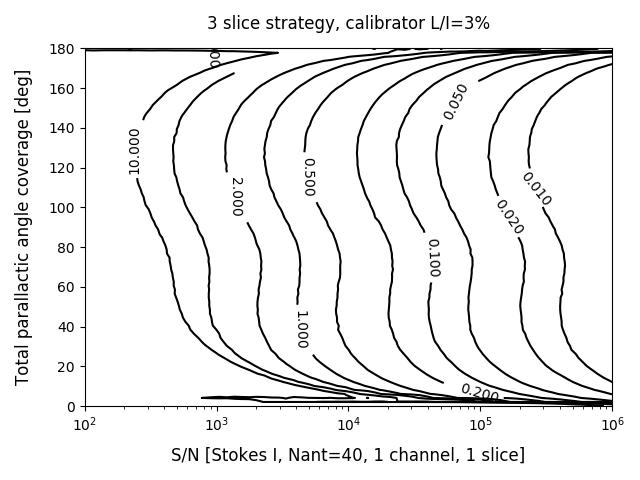}
\includegraphics[clip,width=0.5\textwidth]{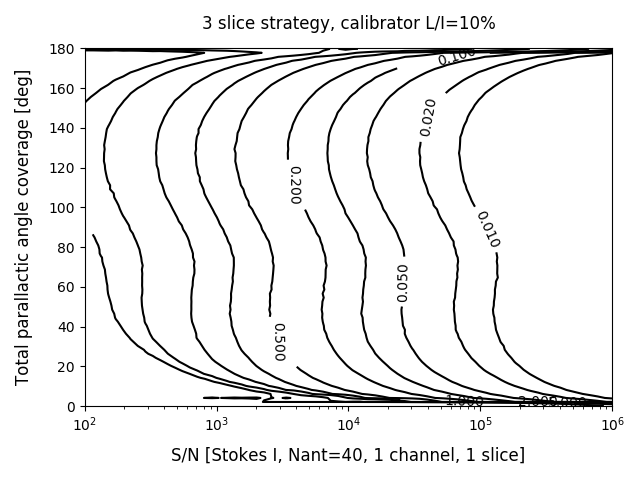}
}
\vspace{0mm}
\centerline{
\includegraphics[clip,width=0.5\textwidth]{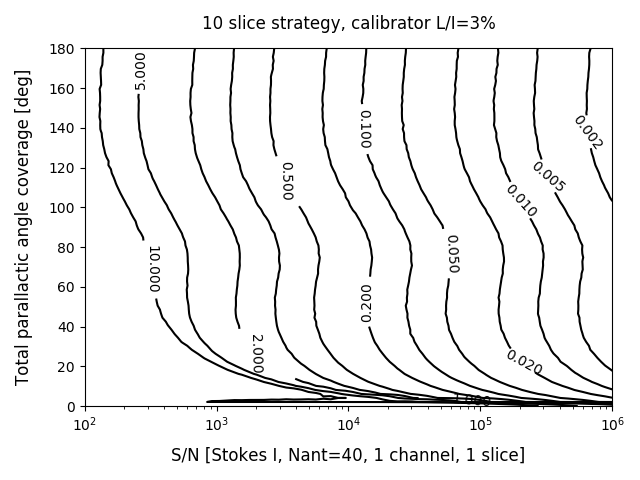}
\includegraphics[clip,width=0.5\textwidth]{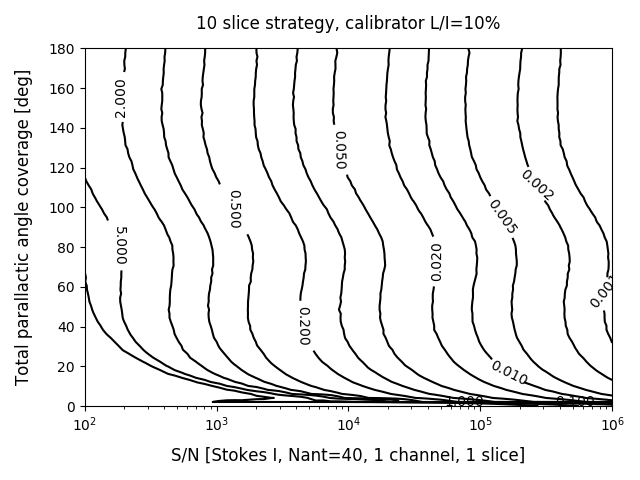}
}
\caption{
Results from simulations showing $95^{th}$ percentile spurious on-axis
fractional linear polarization (percent) for an unpolarized target
arising from different calibration strategies with a linear basis telescope.
The simulations assume an ALMA-like array with 40 antennas.
Top row: 2 slice strategy with polarization known a priori.
Middle row: 3 slice strategy with unknown polarization.
Bottom row: 10 slice strategy with unknown polarization. Panels in the left and right
columns show results obtained using a calibrator with 3\% or 10\% fractional linear
polarization, respectively. Abscissa: Full-array dual-polarization total intensity
signal to noise within 1 spectral channel and 1 slice. Ordinate: Total parallactic
angle coverage; divide by 1 less than the number of slices to get the inter-slice separation.
The distorted contours seen at very small and large parallactic angle coverages in
some panels are the result of coding artifacts and should be ignored.
}
\label{fig:slicesLIN}
\end{figure}

Results for the 2, 3, and 10 slice strategies are displayed in Figure~\ref{fig:slicesLIN}.
The distorted contours seen at very small or very large parallactic angle coverages
in panels where the Stokes vector is unknown are artifacts that can be ignored;
these are the result of simplifications in the code that are localized to these
regions of parameter space (the temporal ordering of points is ignored).
The plots indicate that, in general for a given calibrator,
total parallactic angle coverage of approximately $30^\circ$ is sufficient to maximize
calibration accuracy. Beyond $30^\circ$, additional parallactic angle coverage only
delivers minor improvements. In the 2 slice strategy, separation of slices beyond
approximately $70^\circ$ will lead to degraded calibration solutions and increased
spurious polarization. The spurious polarization levels are found to be smaller
when using a more highly fractionally polarized calibrator, as expected.

\subsubsection{Circular basis}\label{sec:limits:pol:circ}

To calibrate leakages in the circular basis using a polarized source, observations
are required over at least 2 or 3 parallactic angle slices depending on whether the
Stokes vector is known a priori or not, respectively. When unknown a priori, the
Stokes vector needs to be solved for in addition to the d-terms.

Monte Carlo simulations similar to those described for the linear basis were performed
to predict the level of spurious on-axis polarization and absolute position angle error
resulting from a 2 (linearly polarized with Stokes known), 3 (Stokes unknown), and
10 (Stokes unknown) slice strategy. To solve for $\sigma_d$, the code focuses on a
single cross product ($V_{RL}$) and examines how well the d-term for the antenna
and polarization under consideration can be recovered while taking into account all
available $N_a-1$ baselines toward antenna $i$. The simulations assume a VLA-like array
with 27 antennas. Full details are presented in the Appendix, including a web link to
the publicly available simulation code. This section will present the results for
spurious polarization, which is calculated in the code by taking the recovered
$95^{th}$ percentile $\sigma_d$ and performing a conversion using
Equation~\ref{eqn:residL:CF}. Position angle errors will be reported in
Section~\ref{sec:pa:circ}.

\begin{figure}[p]
\centerline{
\includegraphics[clip,width=0.5\textwidth]{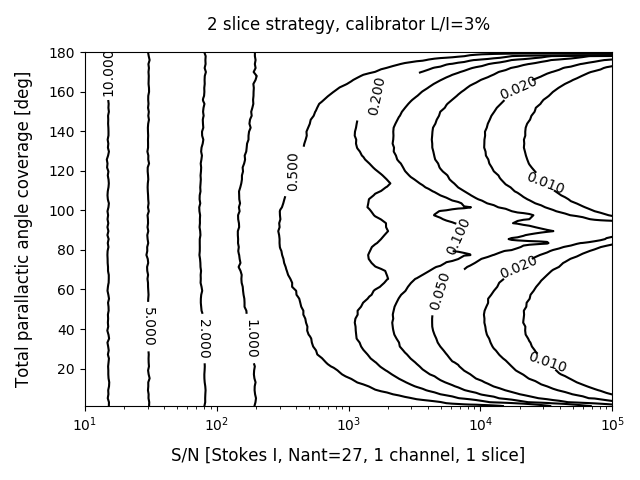}
\includegraphics[clip,width=0.5\textwidth]{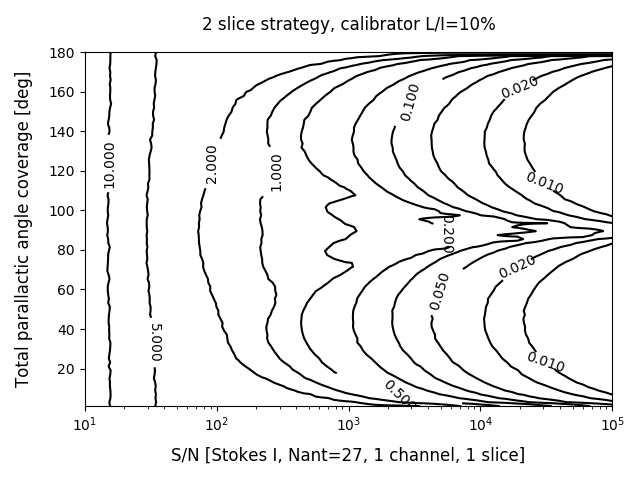}
}
\vspace{2mm}
\centerline{
\includegraphics[clip,width=0.5\textwidth]{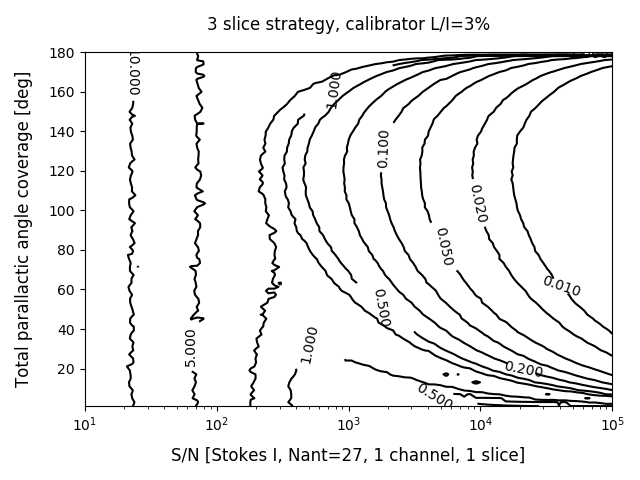}
\includegraphics[clip,width=0.5\textwidth]{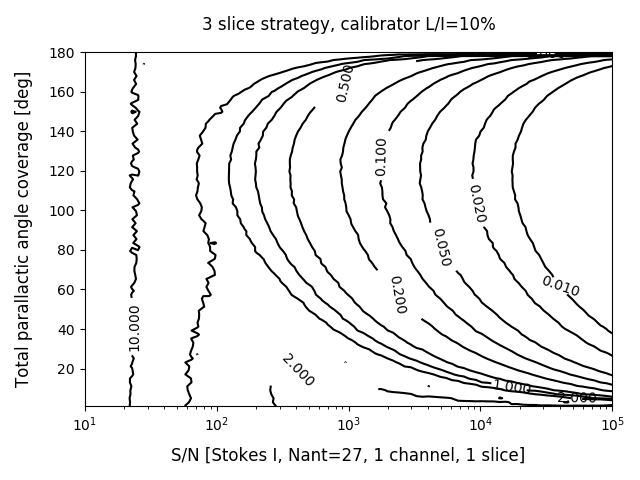}
}
\vspace{2mm}
\centerline{
\includegraphics[clip,width=0.5\textwidth]{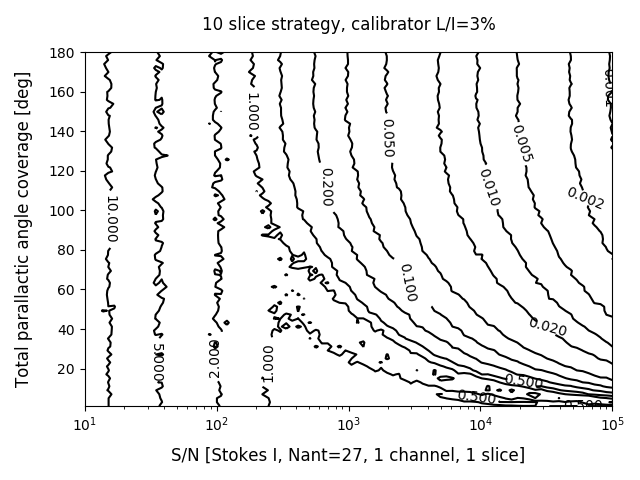}
\includegraphics[clip,width=0.5\textwidth]{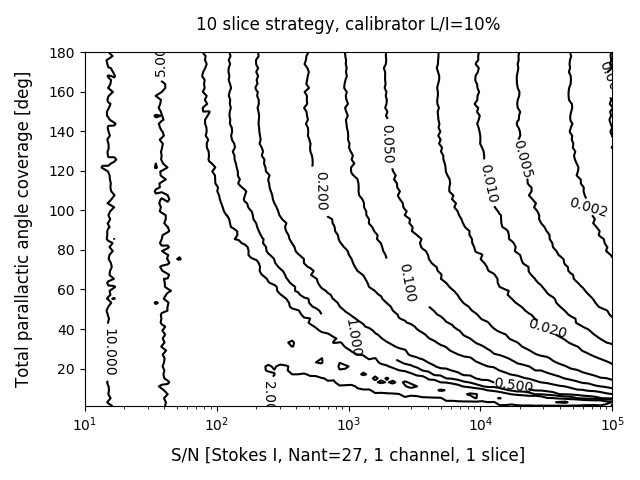}
}
\caption{
Results from simulations showing $95^{th}$ percentile spurious on-axis
fractional linear polarization (percent) for an unpolarized target
arising from different calibration strategies with a circular basis telescope.
The simulations assume a VLA-like array with 27 antennas.
Panel layout and axes are the same as Figure~\ref{fig:slicesLIN}.
The distorted contours seen at small parallactic angle coverages in some panels,
in which a valley is produced, are the result of coding artifacts and should be ignored.
}
\label{fig:slicesCIRC}
\end{figure}

Results are displayed in Figure~\ref{fig:slicesCIRC}.
The distorted contours seen at small parallactic angle coverages in panels where the Stokes
vector is unknown are artifacts that can be ignored; they are the result of simplifications
in the code that are localized to these regions of parameter space (the temporal ordering of
points is ignored). The plots demonstrate that, as expected, a floor is reached at low signal
to noise where no amount of parallactic angle coverage can make up for the dominant randomizing
influence of thermal noise. This is not seen in the linear basis results because the method of
solving for the d-terms is different (see Appendix; linear basis calibration can take advantage
of prior crosshand phase calibration). Note that the displayed signal to noise range differs
between the linear and circular basis plots.

The 2 slice strategy reveals the counter-intuitive result that a calibrator
with larger fractional polarization will, at low signal to noise, result in higher
spurious polarization than a calibrator with low fractional polarization. This is
because the fractional polarization is a fixed known quantity when signal
to noise is defined for total intensity; solving for the origin
of a circle with fixed radius in the presence of thermal noise leads to larger
fractional errors when the radius is larger. Indeed, this trend continues to the
case of unpolarized calibrators; spurious polarization limits are even smaller when
observing an unpolarized calibrator at the equivalent signal to noise. For the 3 and 10
slice strategies, the Stokes vector is not known a-priori, so calibrators with higher
fractional polarization deliver better quality solutions than lower ones, as expected.

The results presented here indicate that, in general, when the Stokes vector of the
leakage calibrator is known a priori, total parallactic angle coverage of approximately
$30^\circ$ is sufficient to maximize calibration accuracy. Additional coverage is
not found to deliver significant improvements. In the 2 slice strategy, separation
of slices beyond approximately $70^\circ$ will lead to degraded calibration solutions
and increased spurious polarization. These match findings for the linear basis presented in
Section~\ref{sec:limits:pol:lin}. However, when the Stokes vector is unknown in
the circular basis, the minimum coverage requirement increases to approximately
$90^\circ$.

\section{Position angle errors}\label{sec:pa}

As with leakage calibration, observational constraints may not always be available to
perform absolute position angle calibration. If absolute position angles are not calibrated,
scientifically useful data may still result, for example if the spectrum of fractional
polarization is of interest, or in particular for the linear basis if position angles
are only partially calibrated. The following sections explore position angle errors
in the linear and circular bases.

\subsection{Linear basis}\label{sec:pa:lin}

In the linear basis, calibration strategies must recover $\operatorname{Re}(d_{X\,ref})$,
the real part of the d-term on the X feed of the gain reference antenna, in order to
provide self-consistent alignment to the assumed sky frame. If relative leakages are
recovered, a systematic contribution to position angle errors will be imposed, given
by the magnitude of $\operatorname{Re}(d_{X\,ref})$. For example,
$\operatorname{Re}(d_{X\,ref})\sim2\%$ implies a systematic position angle error
contribution of $\sim1^\circ$. This relationship can be derived by considering
how the unaccounted degree of freedom associated with the true value of $d_{X\,ref}$
will be absorbed by $\mathcal{Q_\psi}$ and $\mathcal{U_\psi}$ in
Equations~\ref{eqn:vxx}--\ref{eqn:vyy} when relative leakages ($r$) are utilized,
compared with their original values calculated in the presence of absolute leakages ($a$).
The differences are
$\mathcal{Q_\psi}_r - \mathcal{Q_\psi}_a = -2\operatorname{Re}(d_{X\,ref})\,\mathcal{U_\psi}_a$
and
$\mathcal{U_\psi}_r - \mathcal{U_\psi}_a = 2\operatorname{Re}(d_{X\,ref})\,\mathcal{Q_\psi}_a$.
The position angle difference is then $\operatorname{Re}(d_{X\,ref})$ in the small
angle approximation.

The above calculation can also be used to estimate position angle errors resulting
from d-term statistical measurement errors. The resulting worst-case position angle
error is approximately $\sigma_d$, i.e. if all d-terms were made relative to an
offset of this magnitude.

An additional systematic position angle error will arise due to each antenna in the
array exhibiting a non-zero mechanical feed alignment uncertainty about the nominal
alignment (where the nominal alignment is a specified angle between the $X$
feed and the meridian at $\psi=0$). Thus, in general, the standard error in the mean
of feed alignments over the array will be non-zero, leading to an offset between the
assumed and true sky frames. In turn, measured Stokes $\mathcal{Q}$
and $\mathcal{U}$ will be slightly rotated versions of their true values.
Leakage calibration will remain internally consistent for both relative and absolute
cases, accounting for the misaligned feeds. However, the systematic misalignment over
the array will remain uncorrected. External position angle calibration, using a source
with known Stokes vector to adjust $\operatorname{Re}(d_{X\,ref})$ and perform
relative application to all other leakages, is required to account for this offset
and provide true absolute position angle calibration. For statistically independent
feed misalignments over the array (not necessarily true in practice due to
correlated installation procedures), the systematic position angle uncertainty
contribution will be approximately $\phi/\sqrt{N_a}$, where $\phi$ is the characteristic
alignment uncertainty per antenna. To illustrate, this is
{\footnotesize $\lesssim$}~$2^\circ/\sqrt{40}\approx 0.3^\circ$ for ALMA.

Thus, total systematic position angle uncertainty (i.e. frame error) for a target
source in the linear basis is the quadrature sum of up to 3 terms: systematic error
when relative leakages are recovered (i.e. when $d_{X\,ref}$ is artificially
set to complex zero) given by the magnitude of the true $\operatorname{Re}(d_{X\,ref})$,
systematic error from d-term measurement errors, and systematic feed misalignment error.
The relationship is
\begin{equation}
    \theta_{\textrm{syst.}}^{\,2} \approx
    \left[\operatorname{Re}(d_{X\,ref})\right]^2 + \sigma_d^2 + \frac{\phi^2}{N_a} \;\;.
    \label{eqn:pa}
\end{equation}
Total position angle error can then be calculated for a target source by combining
the total systematic error above with statistical error (i.e. in-frame error) associated
with the signal to noise of detection.

In this section, and in the simulation code developed for this work, it will be assumed
that linear basis calibration does {\it not} include absolute position angle calibration
(i.e. the final term in Equation~\ref{eqn:pa} will be included in all calculations). This is
regardless of whether relative or absolute leakages are recovered, or importantly whether a
linearly polarized calibrator with known Stokes vector is available or not. This is consistent
with typical data reduction procedures for major telescopes such as ALMA and the Australia
Telescope Compact Array (ATCA). Results for position angle errors can therefore be
compared directly between the calibration strategies examined in this work. To include
the effect of absolute position angle calibration in any of the results, subtract
$\phi/\sqrt{N_a}$ in quadrature, where $\phi=2^\circ$ is the value assumed in the simulations.
Systematic position angle errors resulting from the use of unpolarized and polarized
calibrators will now be examined.

If an unpolarized calibrator is utilized for polarization calibration, relative
leakages will be recovered. The total systematic position angle error is then estimated
using Equation~\ref{eqn:pa}, including the first term. A worst-case estimate for
the second term is given by $\sigma_d^2=N_a/(2A^2)$ from Equation~\ref{eqn:residpolLF}.

Position angle errors arising from calibration strategies involving a polarized calibrator
were examined using the 1, 2, 3, and 10 slice Monte Carlo simulations introduced in
Section~\ref{sec:limits:pol:lin}. Position angle errors were calculated in a similar
manner to leakages, taking the recovered $95^{th}$ percentile $\sigma_d$ from the
simulations and evaluating the total systematic position angle error using Equation~\ref{eqn:pa}.
The first term in Equation~\ref{eqn:pa} was only included for the 1 slice simulation;
the other calibration strategies recover absolute leakages.

The result from the 1 slice simulation is displayed in Figure~\ref{fig:1sliceknown}.
The asymptotic behavior is due to the final term in Equation~\ref{eqn:pa}. As with
the spurious leakage curve, the overall position angle error curve is practically
indistinguishable from the analytic prediction for an unpolarized calibrator that
may be obtained by combining Equation~\ref{eqn:residpolLF} with Equation~\ref{eqn:pa}.

\begin{figure}[p]
\centerline{
\includegraphics[clip,width=0.5\textwidth]{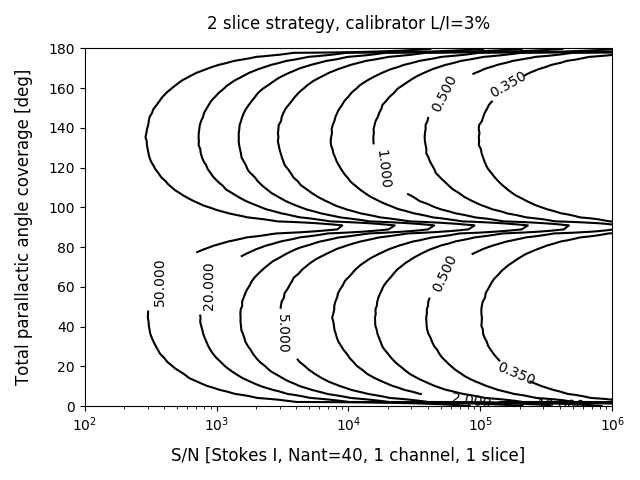}
\includegraphics[clip,width=0.5\textwidth]{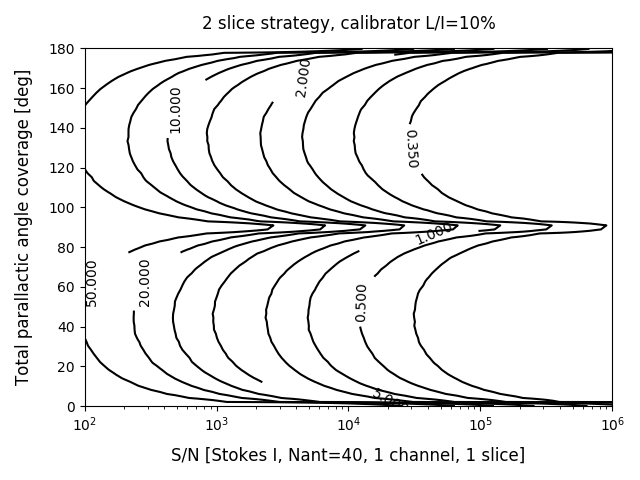}
}
\vspace{2mm}
\centerline{
\includegraphics[clip,width=0.5\textwidth]{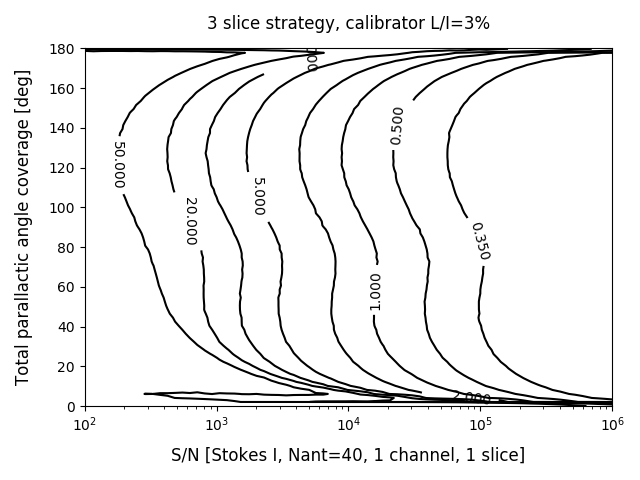}
\includegraphics[clip,width=0.5\textwidth]{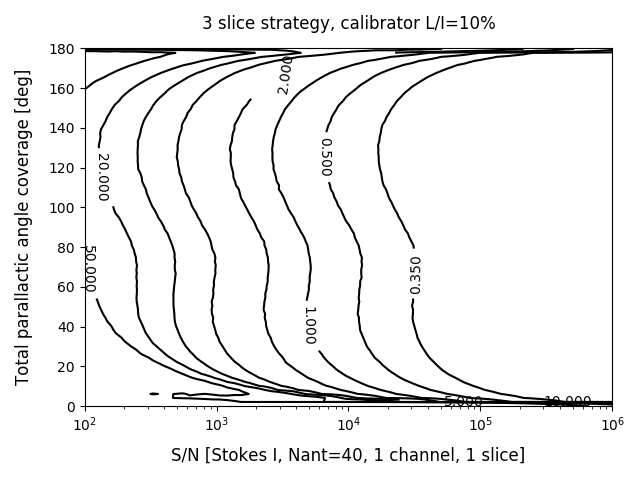}
}
\vspace{2mm}
\centerline{
\includegraphics[clip,width=0.5\textwidth]{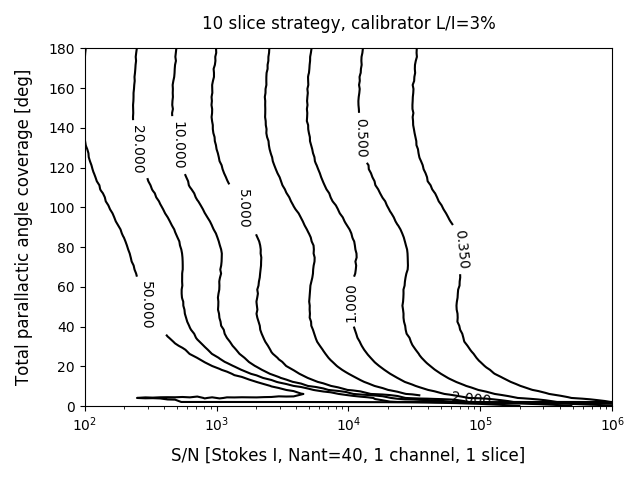}
\includegraphics[clip,width=0.5\textwidth]{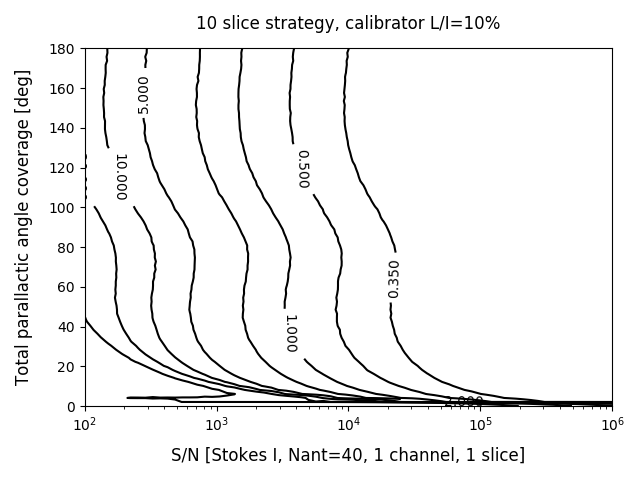}
}
\caption{Results from simulations showing $95^{th}$ percentile systematic
position angle error (degrees) arising from different calibration strategies
with a linear basis telescope. Panel layout and axes are the same as Figure~\ref{fig:slicesLIN}.
The distorted contours seen at very small and large parallactic angle coverages in
some panels are the result of coding artifacts and should be ignored.
}
\label{fig:slicesLINpa}
\end{figure}

Results from the 2, 3, and 10 slice strategies are displayed in Figure~\ref{fig:slicesLINpa}.
Similar conclusions regarding parallactic angle coverage may be drawn as for leakages
described in Section~\ref{sec:limits:pol:lin}.

\subsection{Circular basis}\label{sec:pa:circ}

Position angle calibration in the circular basis is tied to crosshand phase calibration.
This requires a polarized calibrator.

\begin{figure}
\centerline{\includegraphics[clip,width=0.5\textwidth]{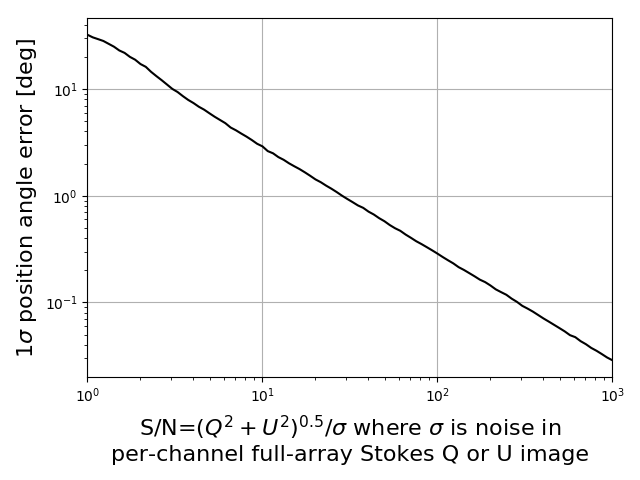}}
\caption{
Result from simulation showing systematic position angle error for calibration
with a circular basis telescope. Abscissa: Full-array dual-polarization linear polarization
signal to noise within 1 spectral channel. Ordinate: Standard error in systematic position
angle error.
}
\label{fig:calibpa}
\end{figure}

To examine position angle errors resulting from crosshand phase calibration, a Monte
Carlo simulation was performed based on Equation~\ref{eqn:rlphase}. The result is
presented in Figure~\ref{fig:calibpa}.
To illustrate interpretation of this figure, consider position angle calibration with the
VLA within a 2~MHz channel at 3~GHz using 3C48 ($\sim10$~Jy, $\sim2$\% fractional linear polarization).
A signal to noise ratio in excess of 300 is required to limit systematic position angle uncertainty
to within $0.1^\circ$. This translates to an on-source time approaching 4~min. For $0.3^\circ$
uncertainty, the required on-source time is $\sim30$~sec.

\section{Conclusions}\label{sec:conclusions}

The mathematical framework for describing interferometric radio polarimetry does not readily
permit quantitative calculation of post-calibration residuals for specific observational
calibration strategies. This work has bridged this gap through the presentation of analytic
derivations and results from Monte Carlo simulations. In general, worst-case errors have been
assumed. Thus, in practice, residual leakage and position angle errors are likely to be smaller.

This work has focused on arrays that comprise alt-az antennas with common feeds over
which parallactic angle is approximately uniform, such as ALMA and the VLA. The simulation
code has been made publicly available to support potential extension, for example to investigate
mixed basis arrays \citep[e.g.][]{2016A&A...587A.143M}, VLBI, calibration strategies using resolved
polarization calibrators, or a more detailed examination of circular polarimetry.

This work was motivated by the need to implement automated polarization data reduction capabilities
within the CASA integrated pipelines for ALMA\footnote{\url{https://almascience.nrao.edu/documents-and-tools/}}
and the VLA\footnote{\url{https://science.nrao.edu/facilities/vla/data-processing/pipeline}}.
As a result, this work forms part of ALMA Memo 603, equivalently referenced as EVLA Memo 201
\citep{halesmemo}. This memo contains extensive additional material, including a general classification
system for polarization calibrators, and detailed step by step procedures for performing a
suite of polarimetric calibration strategies in the linear and circular bases.

\acknowledgments

I thank the following for insightful discussions:
George Moellenbrock, Bob Sault, Brian Kent, Lindsey Davis,
Vincent Geers, Kumar Golap, Jeff Kern, Joe Masters, and Claire Chandler.
I thank the ALMA Integrated Science Team for reviewing this document
under the context of CASA automated pipeline design. I thank the anonymous
referee for their thoughtful review which led to the improvement of this
paper. The National Radio Astronomy Observatory is a facility of the National
Science Foundation operated under cooperative agreement by Associated Universities,
Inc. This project has received funding from the European Union's Horizon 2020
research and innovation programme under the Marie Sk{\l}odowska-Curie
grant agreement No 705332.

\software{Simulation code \citep{hales2017},
          CASA \citep{2007ASPC..376..127M}}

\appendix

\section{Residual on-axis instrumental leakage}

Section~\ref{sec:limits} presented equations to predict the level of spurious
on-axis polarization that will be observed for an intrinsically unpolarized
target source following the application of imperfect d-term calibration solutions.
Using these relationships, Section~\ref{sec:limits:unpol} presented equations
to predict the d-term measurement errors, and in turn the level of spurious
polarization, that will result following leakage calibration when using a
polarized calibrator that is assumed to be unpolarized. Similarly,
Section~\ref{sec:limits:pol} presented results from simulations in
which d-term measurement errors, and ultimately spurious polarization signatures,
were predicted empirically for calibration schemes involving observation
of a polarized calibrator over a range of parallactic angles (slices).
%Sections~\ref{sec:limits:unpol} and \ref{sec:limits:pol} also presented
%estimated position angle errors.
Derivations for all equations, and details of the simulations, are presented
below for the circular and linear feed bases.

%B1: convert sigma d into spurious L (Section~\ref{sec:limits})
%B2: unpolarized source (Section~\ref{sec:limits:unpol}), fig 2
%B3: polarized source 2/3/10 slice, lin (figs 3,4) and circ bases (figs 5,6) (Section~\ref{sec:limits:pol})

\subsection{Circular basis}

Stokes $\mathcal{Q}$ is formed by
\begin{equation}
    \mathcal{Q} = 0.5 \Re\left[ \left<e^{+i2\psi}\,V_{RL}\right> +
                                \left<e^{-i2\psi}\,V_{LR}\right> \right] \,.
\end{equation}
If d-terms are recovered with measurement errors $\Delta d$ (statistical or systematic
in origin), an unpolarized target will be observed with spurious fractional polarization
\begin{equation}
    \frac{\mathcal{Q}_{\epsilon}}{\mathcal{I}} = 0.5 \Re\left[
               \left<e^{+i2\psi}\,\left(\Delta d_{Ri} + \Delta d_{Lj}^{\,*}\right)\right> +
               \left<e^{-i2\psi}\,\left(\Delta d_{Li} + \Delta d_{Rj}^{\,*}\right)\right> \right] \,.
\end{equation}
The worst-case spurious polarization will therefore occur for a target
observed with limited parallactic angle coverage. If the science target is
integrated over a wide range in parallactic angle, then the level of spurious
polarization predicted in the following should be treated as an upper limit. Taking
the worst-case scenario of approximately constant parallactic angle, and
noting that there are only $N_a$ independent d-terms per polarization over the array,
the relationship above can be rewritten in a statistical sense as
\begin{equation}\label{eqn:appB:circ:qspur}
    \frac{\mathcal{Q}_{\epsilon}}{\mathcal{I}} \approx \Re\!\bigg[
                  \, \frac{1}{N_a} \sum^{N_a} \Delta d_{Ri} + \Delta d_{Li} \, \bigg] \; .
\end{equation}
For characteristic d-term modulus error $\sigma_d$, the variance in
$\Re\left[\Delta d\right]$ is given by $\sigma_d^2/2$. The variance in
Equation~\ref{eqn:appB:circ:qspur} is then estimated as
\begin{equation}
    \textrm{var}\left(\frac{\mathcal{Q}_{\epsilon}}{\mathcal{I}}\right) \approx
     \frac{\sigma_d^2}{N_a} \; .
\end{equation}
Similar analysis for fractional $\mathcal{U}_{\epsilon}$ yields the same result.
The predicted level of spurious on-axis fractional linear polarization is
then Rayleigh distributed with mean
\begin{equation}\label{eqn:appB:circ:lspur}
    \frac{\mathcal{L}_{\epsilon}}{\mathcal{I}} \approx \sqrt{\frac{\pi}{2 N_a}}\,\sigma_d \; .
\end{equation}
No spurious circular polarization is predicted ($\mathcal{V}_{\epsilon}=0$) because its
evaluation does not include any leakage products with total intensity (to first order).
The results above are presented in Section~\ref{sec:limits}.

If a polarized calibrator is assumed to be unpolarized for leakage calibration,
any true (linear) polarization will lead to corruption of the measured leakages. The difference
between observed and true cross hand visibilities for a single baseline is given by
\begin{equation}
    \Delta V_{RL} = \mathcal{I}\left(\Delta d_{Ri} + \Delta d_{Lj}^{\,*}\right) -
                    (\mathcal{Q_{\rm{true}}} + i\,\mathcal{U_{\rm{true}}})e^{-i2\psi} \; .
\end{equation}
Note that non-zero $\mathcal{V_{\rm{true}}}$ will not affect relative leakages that
are calculated using only cross hand data (to first order). The equation above is
effectively constrained by $N_a-1$ baselines toward antenna $i$, in which case
\begin{equation}\label{eqn:appB:circ:unpol}
    \Delta d_{Ri} + \frac{1}{N_a-1} \sum^{N_a-1} \Delta d_{Lj}^{\,*} =
    \frac{\mathcal{Q_{\rm{true}}} + i\,\mathcal{U_{\rm{true}}}}{\mathcal{I}}e^{-i2\psi} +
    \frac{1}{N_a-1} \sum^{N_a-1} \frac{\Delta V_{RL}}{\mathcal{I}} \; .
\end{equation}
In this construction, the leakages will soak up the source polarization, leaving the $\Delta V_{RL}$
term consisting of only thermal noise. As a result, its average in the right side
of the equation can be represented by a vector with characteristic magnitude $\sqrt{N_a}/A$, where
$A$ is the full-array dual-polarization total intensity signal to noise of the calibrator within
the single spectral channel of interest.  The first term in the left side of the equation
has characteristic magnitude $\sigma_d$. The importance of the next term, containing the
average over d-terms, depends on whether the d-terms are correlated between antennas or not.
When random errors dominate over systematics from the true source polarization (e.g. for small $A$),
the recovered d-term errors will be effectively uncorrelated. In this case, the term can be
viewed as a vector-averaged sample of ($\sigma_d$-scale) error vectors, in which case its
contribution will be negligible\footnote{To demonstrate, consider the variance for a sample of
unit vectors with random orientations projected along a 1D axis. This is given by $0.5/(N_a-1)$.
The standard error for the left side of the equation can therefore be approximated by
$\sigma_d \sqrt{1+0.5/(N_a-1)}$. This indicates a negligible difference of $<12\%$ for $N_a > 2$.}
and can be ignored. When source polarization systematics dominate (e.g. for large $A$),
the d-terms will be correlated and the average cannot be ignored. This can be crudely
accommodated by replacing $\Delta d_{Lj}^{\,*}$ with $\Delta d_{Ri}$, in which case the
left side of Equation~\ref{eqn:appB:circ:unpol} can be approximated by $2\Delta d_{Ri}$.
Thus, the estimated $\sigma_d$ will be half of the value recovered when assuming
uncorrelated d-terms. Given the simplistic nature of this calculation, the
larger estimate for $\sigma_d$ will be adopted; its estimated value presented below
should therefore be treated as an upper limit.

By noting that contributions to $\sigma_d$ on the right side of Equation~\ref{eqn:appB:circ:unpol}
represent projections onto a 1D vector given by the true d-term (requiring adjustment to variances
by factor $1/2$), and by treating the true polarization as a DC offset with magnitude
$\mathcal{L_{\rm{true}}}$, the d-term modulus error can be estimated in rms-fashion as
\begin{equation}\label{eqn:appB:circ:unpol:d}
    \sigma_d \approx \sqrt{\frac{1}{2}\left(\left[\frac{\mathcal{L_\textrm{true}}}{\mathcal{I}}\right]^2 +
        \frac{N_a}{A^2}\right)} \; .
\end{equation}
The resulting estimate for spurious fractional linear polarization is then obtained using
Equation~\ref{eqn:appB:circ:lspur}, giving
\begin{equation}\label{eqn:appB:circ:unpol:l}
    \frac{\mathcal{L}_{\epsilon}}{\mathcal{I}} \approx
        \sqrt{\frac{\pi}{4 N_a} \left(\left[\frac{\mathcal{L_\textrm{true}}}{\mathcal{I}}\right]^2 + 
        \frac{N_a}{A^2}\right)} \; .
\end{equation}
These results are reported in Section~\ref{sec:limits:unpol}. Note that if the leakage
calibrator is observed over a wide range in parallactic angle (atypical for an assumed
unpolarized calibrator), then the predicted spurious polarization should be treated as
an upper limit (in addition to the motivation described earlier).

Figure~\ref{fig:slicesCIRC} presents estimates of spurious polarization for calibration
strategies involving parallactic angle coverage of a polarized leakage calibrator. To
obtain these results, a Monte Carlo simulation code was developed to estimate $\sigma_d$
and perform conversion using Equation~\ref{eqn:appB:circ:lspur}. The code focuses on the
theoretical aspects discussed in this document by approximating the behaviour of the
generalized solvers that exist within software such as CASA, as described below. Full
CASA-based (or other package) simulations using mock or real data were not considered
for this work due to the potential for introducing a host of unwanted systematics,
which could readily bias interpretation of the fundamental attributes under
investigation. The simulation code is publicly available at
\url{https://github.com/chrishales/polcalsims} \citep{hales2017}.

To estimate $\sigma_d$ for the calibration schemes examined, the code performs
Monte Carlo sampling and examines the distribution of errors recovered when attempting
to solve for the d-term for a single polarization on a single antenna. To do this,
the code focuses on a single cross product (e.g. $V_{RL}$) and examines how well
the d-term under consideration can be recovered while taking into account all
available $N_a-1$ baselines toward antenna $i$. The relevant relationship is
given by Equation~\ref{eqn:appB:circ:unpol}, with the sum over $\Delta d_{Lj}^{\,*}$
assumed to be negligible. The true source polarization is injected with the
appropriate thermal noise at slices that are, for simplicity, spaced equally
over the total parallactic angle span under consideration.

Calibration strategies involving a polarized calibrator with unknown
Stokes vector require at least 3 statistically independent slices to
solve for the d-term (error) as well as Stokes $\mathcal{Q}$ and $\mathcal{U}$.
Geometrically, this can be viewed as the need for 3 points to solve for
the unknown origin and radius of a circle (see Fig.~\ref{fig:geomRL}).
When the Stokes vector is known a priori, only two slices are required to
locate the origin (the origin degeneracy is broken by the known sense of
rotation between the slices). For simplicity, the simulation code does not
take into account the sense of rotation between points. As a result, the
portion of parameter space containing observations at modest signal to noise
ratios over small total parallactic angle ranges displays much noisier
solutions than those likely to be recovered in production code. This
effect is not significant; results throughout the remaining parameter
space are not affected.

The code recovers the distribution of d-term errors for each sampled point
in the signal to noise and parallactic angle coverage parameter space. Rather
than reporting the mean of this distribution to represent $\sigma_d$, the
code reports the $95^{th}$ percentile in order to better accommodate the
slightly non-Gaussian nature of the results in a conservative manner.
This is consistent with the comments earlier to interpret results as
upper limits.

\subsection{Linear basis}

Assuming perfect crosshand phase measurement, $\mathcal{U_\psi}$ is formed by
\begin{equation}
    \mathcal{\mathcal{U_\psi}} = 0.5 \Re\left[ \left<V_{XY}\right> +
                                               \left<V_{YX}\right> \right] \,.
\end{equation}
The presence of d-term measurement errors will cause an unpolarized target to exhibit
spurious fractional linear polarization, described statistically as
\begin{equation}\label{eqn:appB:lin:qspur}
    \frac{\mathcal{U_{\psi,\epsilon}}}{\mathcal{I}} \approx \Re\!\bigg[
                  \, \frac{1}{N_a} \sum^{N_a} \Delta d_{Xi} + \Delta d_{Yi} \, \bigg] \; .
\end{equation}
The worst-case spurious polarization will occur for a target observed with limited
parallactic angle coverage. Assuming the worst-case scenario of approximately constant
parallactic angle, the variance in Equation~\ref{eqn:appB:lin:qspur} is then estimated as
\begin{equation}
    \textrm{var}\left(\frac{\mathcal{U_{\psi,\epsilon}}}{\mathcal{I}}\right) \approx
     \frac{\sigma_d^2}{N_a} \; .
\end{equation}
Similar analysis for fractional $\mathcal{V}_{\epsilon}$ yields the same result.
No spurious $\mathcal{Q_{\psi}}$ will be produced because its evaluation does not
include any leakage products with total intensity. As a result, the predicted level
of spurious on-axis fractional linear or circular polarization is given by
\begin{equation}\label{eqn:appB:lin:lv}
    \frac{\mathcal{L}_{\epsilon}}{\mathcal{I}} \approx
    \frac{\mathcal{V}_{\epsilon}}{\mathcal{I}} \approx \frac{\sigma_d}{\sqrt{N_a}} \; .
\end{equation}
The predicted level of spurious fractional elliptical polarization is then Rayleigh
distributed with mean
\begin{equation}\label{eqn:appB:lin:p}
    \frac{\mathcal{P}_{\epsilon}}{\mathcal{I}} \approx \sqrt{\frac{\pi}{2 N_a}}\,\sigma_d \; .
\end{equation}
These results are presented in Section~\ref{sec:limits}.

The results presented in Section~\ref{sec:limits:unpol} regarding calibration with
a polarized yet assumed-unpolarized calibrator can be derived in the same way as
presented earlier for the circular feed basis, but replacing $\mathcal{L}^2_\textrm{true}$
with $\mathcal{U}^2_{\psi ,\textrm{true}}+\mathcal{V}^2_\textrm{true}$ in
Equation~\ref{eqn:appB:circ:unpol:d}. For the linear basis derivation here, it will be
assumed that the product of $\mathcal{Q}_{\psi ,\textrm{true}}$ with leakages in the
cross hand visibilities is always negligible. This will not always be true in practice,
but in such cases the contribution from thermal noise ($A$) is likely to dominate.
The resulting estimates of spurious fractional linear or circular polarization
are then obtained using Equation~\ref{eqn:appB:lin:lv}, giving
\begin{equation}
    \frac{\mathcal{L}_{\epsilon}}{\mathcal{I}} \approx
    \frac{\mathcal{V}_{\epsilon}}{\mathcal{I}} \approx
        \sqrt{\frac{1}{2} \left(\left[\frac{\mathcal{U}_{\psi ,\textrm{true}}}{\mathcal{I}}\right]^2 + 
                                \left[\frac{\mathcal{V}_\textrm{true}}{\mathcal{I}}\right]^2 +
                                \frac{N_a}{A^2}\right)} \; .
\end{equation}
The estimate of spurious fractional elliptical polarization is obtained using
Equation~\ref{eqn:appB:lin:p}, giving
\begin{equation}
    \frac{\mathcal{P}_{\epsilon}}{\mathcal{I}} \approx
        \sqrt{\frac{\pi}{4 N_a} \left(\left[\frac{\mathcal{U}_{\psi ,\textrm{true}}}{\mathcal{I}}\right]^2 + 
                                      \left[\frac{\mathcal{V}_\textrm{true}}{\mathcal{I}}\right]^2 +
                                      \frac{N_a}{A^2}\right)} \; .
\end{equation}

Figure~\ref{fig:slicesLIN} presents estimates of
spurious polarization for calibration strategies involving parallactic angle coverage
of a polarized leakage calibrator. To obtain these results, simulation code was developed
with similar characteristics to those described earlier for the circular feed basis.
Differences are described below. The simulation code is publicly available at
\url{https://github.com/chrishales/polcalsims} \citep{hales2017}.

The code focuses on a single cross product (e.g. $V_{XY}$) and examines how well
the d-term for the antenna and polarization under consideration can be recovered
while taking into account all available $N_a-1$ baselines toward antenna $i$.
Unlike in the circular basis, measurement of the crosshand phase is required
here prior to solving for leakages. The linear basis simulation code therefore
takes into account crosshand phase measurement errors due to thermal noise when
calculating the d-term measurement errors. The code does not account for errors
in crosshand phase measurement due to the as-yet unknown leakages, which are
assumed to be zero for this calculation (such errors are typically negligible
in the baseline-averaged crosshand phase solve, minimizing the need for iteration).
The code assumes that Stokes $\mathcal{V}$ is zero for all calibrators. The
relevant equation is then a modified version of Equation~\ref{eqn:vxy}, 
\begin{equation}\label{eqn:appB:lin:pol}
    \frac{1}{N_a-1} \sum^{N_a-1} \frac{V_{XY}}{\mathcal{I}} =
    \frac{\mathcal{U_\psi}}{\mathcal{I}}e^{i\rho} +
    \left( 1-\frac{\mathcal{Q_\psi}}{\mathcal{I}} \right)e^{i\rho}\,d_{Xi} +
    \left( 1-\frac{\mathcal{Q_\psi}}{\mathcal{I}} \right)
    \frac{e^{i\rho}}{N_a-1}\sum^{N_a-1} d_{Yj}^{\,*} + 
    \frac{1}{N_a-1} \sum^{N_a-1} \frac{\sigma_{V_{XY}}}{\mathcal{I}}
\end{equation}
in which thermal noise in $V_{XY}$ is explicitly included, denoted by $\sigma_{V_{XY}}$.
For all multi-slice observing strategies, the simulation code assigns each of the d-terms
that appear in the equation above with a user-defined characteristic amplitude and
random phase. The error in recovering the input $d_{Xi}$ is then ultimately recorded.

For simplicity, the code assumes that the first observed slice for each
calibration strategy is at zero parallactic angle, and that the calibrator's
position angle is $45^\circ$. These initial conditions should generate generally
representative results for the 1 and 2 slice strategies, where the calibrator's
Stokes vector is known a priori and may therefore be targeted appropriately by
observers. However, note that the initial conditions above (or any others)
cannot fully represent all possible observing configurations for the 3
and 10 slice strategies, where the Stokes vector is unknown a priori. It is of
course possible that rare specific configurations of these strategies could
produce significantly different results than presented. It is worth noting here
that when users in the linear feed basis are advised to maximize parallactic
angle coverage, this really means they should maximise coverage for
$\mathcal{U_\psi}$ (i.e. maximum arc length traced along ellipse for a cross hand
visibility). For a calibrator with unknown Stokes vector observed
over as few as 3 slices, the difference in rare circumstances could be noticeable.

For the 1 slice strategy, the simulation code measures crosshand phase error
by solving for a position angle in the noisy frame indicated by
Equation~\ref{eqn:chp:lin} (i.e. {\tt Xf} solve in CASA terminology).
For the 2 slice strategy, the code performs this step using the slice
with maximum $\mathcal{U_\psi}$ (known a priori). For the 3 and 10 slice
strategies, the code measures crosshand phase error by solving for a linear
slope with unconstrained offset, followed by a least squares fit to measure
Stokes $\mathcal{Q}$ and $\mathcal{U}$ along this slope given the noisy
observed variations of $\mathcal{U_\psi}$ (i.e. {\tt XYf+QUf} solve in CASA
terminology).

For the multi-slice strategies, the code then performs a least squares fit
to solve for two parameters in Equation~\ref{eqn:appB:lin:pol}: the observed
$d_{Xi}$ and $\sum d_{Yj}^{\,*}$. The latter is not needed for further analysis.
The former is compared with the input value to compute the d-term error
for the Monte Carlo sample under consideration, followed by conversion to
spurious polarization using Equation~\ref{eqn:appB:lin:lv}. For the 1 slice
strategy, the d-term error can be calculated more easily as the offset from
the noisy measurement of the known Stokes vector.

\end{document}